\documentclass[journal,9pt]{IEEEtran}

\usepackage{cite}
\usepackage{amsmath,amssymb,amsfonts}
\usepackage{algorithmic}
\usepackage{graphicx}
\usepackage{textcomp}

\begin{document}

\title{A Method for Localization of Cellular Users from Call Detail Records}
\author{
Steven W.\ Ellingson\thanks{S. W. Ellingson is with the Bradley Department of Electrical and Computer Engineering, Virginia Tech, Blacksburg, VA, 24061 USA e-mail: ellingson@vt.edu.}
\IEEEmembership{Senior Member, IEEE} 
}

\maketitle

\begin{abstract}
A common problem in justice applications is localization of a user of a cellular network using a call detail record (CDR), which typically reveals only the base station and sector to which the user was connected.
This precludes precise estimation of location. 
Instead, one is limited to estimating a region of plausible locations (RPL) using static information such as sector antenna orientation, beamwidth, and locations of nearby base stations. 
In this paper, we propose a method for RPL estimation in which the shape bounding the RPL is derived from a model of the antenna pattern via the Friis Transmission Equation, 
and the size of the RPL is determined by mean distance to nearby base stations.
The performance of the proposed method is evaluated by ``best server'' analysis of measurements acquired from drive testing in the vicinity of Winter Garden, Florida, observing three 700~MHz-band LTE cellular networks serving this area. 
Of the 16 sectors evaluated, the aggregate error rate (i.e., fraction of users located outside the RPL estimated for the associated sector) is found to be 1.3\%, with worst per-sector error rate of about 13.3\% and error rates below 1.8\% for 13 of the 16 sectors.
The principal difficulty is shown to be estimation of RPL size, which entails a tradeoff between minimizing RPL area (yielding the ``tightest'' localization) and minimizing error rate. 
\end{abstract}

\section{Introduction}
\label{sIntro}

A long standing problem in cellular telecommunications is determining the location of a mobile station (MS; i.e., a user) given information obtained from the cellular infrastructure.  
Technical solutions involving measurements of time difference of arrival, direction of arrival, and RF fingerprinting are generally well known and have been extensively studied; see e.g. \cite{P+17} and references therein.
This paper addresses a particular form of the problem which has received relatively little attention: Localization of an MS using only the information provided in a call detail record (CDR).
A CDR is a record of a connection involving the MS, indicating the particular sector serving the MS.
Note here the term ``sector'' refers to a particular subset of equipment within a base station (BS), and not the geographical region which this equipment is intended to serve.
Whereas the CDR may uniquely identify a particular antenna at a particular BS, the associated geographical footprint of the sector is usually not provided and is typically not precisely known. 

CDRs are the usual response when the operator of a cellular system is served legal process
for MS location by law enforcement.
In the well-known ``Cell ID'' (CID) method (again, see e.g. \cite{P+17}), the MS position is estimated to be the location of the BS to which the MS is connected, with uncertainty understood to be on the order of the spacing between BSs.
To do better, additional technical detail is required.
However, various technical, business, and legal constraints preclude access to information that could be used to estimate position. 
Instead, one is limited to estimating a \emph{region} in which the MS is located.
In principle, this region is the geographical footprint served by the sector.
However, this footprint is difficult to define for various reasons including hand off behavior and complex propagation conditions. 
While operators may have relevant information about this footprint -- perhaps as part of the network design process, or through field testing -- they are typically not required to provide this information.
Moreover, an operator may be disincentivized to provide this information due to concerns about accuracy, liability, and the possibility of exploitation of this information by business competitors. 

In contrast, operators typically \emph{are} willing to provide antenna azimuthal orientations, azimuthal beamwidths, and frequencies.
Additional infrastructure information such as antenna height is sometimes, but not consistently, available.
This information can be used to estimate a region within which the MS could plausibly be located; we refer to this as a ``region of plausible locations'' (RPL). 
The RPL is a natural way to define localization in law enforcement and criminal defense applications because identification of regions where the MS \emph{could} versus \emph{could not} be plausibly located may be more useful than either an estimate of position or an estimate of the sector geographic footprint based on undisclosed or unverifiable technical information.
For the purposes of this paper, RPL-based localization is defined as the estimation of the boundary of the smallest RPL with certainty approaching 100\%, using only information which is consistently and reliably provided by operators, namely the azimuthal orientation and beamwidth of the antenna associated with the sector identified in a CDR, and BS locations.  

Aspects of the RPL estimation problem are addressed in detail in \cite{JC22} and \cite{JC22c}.
The principal difficulty in estimating the RPL is the presence of other BSs, and in particular the fact that it is difficult to predict boundaries at which the MS will hand off between sectors. 
This has led courts to deem inadmissible RPL estimates that depend on ``granulization;'' that is, estimation of hand off boundaries (see e.g., \cite{USvEvans}).
This has left law enforcement 
practitioners in the awkward position of having to determine RPLs with definite boundaries using methods that do not explicitly consider hand off between BSs (see e.g. discussion in \cite{JC22} citing \cite{CvCM1} and \cite{CvCM2}). 
 
The principal contribution of this paper is a method for CDR-based localization in which the RPL \emph{shape} is derived from a model of the antenna pattern via the Friis Transmision Equation, using a well-established model for terrestrial radio path loss that is applicable to outdoor macrocells at UHF frequencies.
This shape is then scaled to account for BS spatial density, which bypasses the problem of estimating sector hand off boundaries. 
To the best of our knowledge, this is a novel method for RPL estimation.
The efficacy of the method is demonstrated using drive test measurements.
While insufficient information exists for a meaningful quantitative comparison of performance to other methods addressed in the previous paragraph, the method and results reported here are suitable for future ``apples-to-apples'' comparisons to both (1) the proposed method employed in other scenarios, and (2) other methods when results are reported in sufficient quantitative detail.

This paper is organized as follows:
The RPL estimation method is developed in Section~\ref{sREA}. 
The method requires a model for the azimuthal directivity pattern of the antenna, which is provided in Section~\ref{sRSM}.
In Section~\ref{sFE}, the efficacy of the method is demonstrated by field experiment, with results summarized in Section~\ref{ssSR}.
Conclusions are summarized in Section~\ref{sConc}.

\section{RPL Estimation Method}
\label{sREA}

According to the Friis Transmission Equation (see e.g., \cite{RSE-R1E}), the power $P_R$ received by an MS is 
\begin{equation}
P_R = P_T ~ G_T ~ L^{-1}(r) ~ G_R
\end{equation}
where
$P_T$ is transmit power,
$G_T$ is the gain of the sector antenna in the direction of the MS,
$L(r)$ is path loss at distance $r$ between the sector antenna and the MS, and
$G_R$ is the antenna gain of the MS in the direction of the BS.
Regardless of the location of other BSs, the distance to the RPL boundary must be less than or equal to the distance $r=r_{max}$ at which $P_R=P_{R,min}$, the sensitivity of the MS.
Thus,
\begin{equation}
P_{R,min} = P_T ~ G_T ~ L^{-1}(r_{max}) ~ G_R
\end{equation}

To solve for $r_{max}$, an expression for path loss is needed. 
A model that captures the relevant features of terrestrial propagation is the ``breakpoint power law'' model (see e.g. \cite{RSE-R1E}), represented by the simple expression:
\begin{equation}
L^{-1}(r) = \left( \frac{\lambda}{4\pi r_b} \right)^2 \left(\frac{r_b}{r}\right)^n
\label{eBPL}
\end{equation} 
where 
$n$ is the \emph{path loss exponent}, with $n=2$ (free space propagation) for $r\le r_b$, and $n\ge 2$ (typically 3--6) otherwise, and
$r_b$ is the \emph{breakpoint distance} 
\begin{equation}
r_b = \frac{4\pi}{\lambda}h_{BS} h_{MS}
\end{equation}
where $h_{BS}$ and $h_{MS}$ are the heights above ground of the sector and MS antennas, respectively.
Making the substitution,
\begin{equation}
P_{R,min} = P_T ~ G_T ~ \left( \frac{\lambda}{4\pi r_b} \right)^2 \left(\frac{r_b}{r_{max}}\right)^n ~ G_R
\end{equation}
Solving for $r_{max}$, we obtain:
\begin{equation}
r_{max} = \left(\frac{\lambda}{4\pi}\right)^{2/n} 
                \left(\frac{P_T G_T G_R}{P_{R,min}} \right)^{1/n} 
                r_b^{1-2/n} 
\label{eRDCL}
\end{equation}

It is not normally possible to calculate $r_{max}$ in an CDR-based localization problem.  
Among the many difficulties is that $P_T$ is typically unknown and varies dynamically due to power control, 
the received power may differ from the mean power due to local shadowing and diffraction,
$G_R/P_{R,min}$ varies over orders of magnitude depending on the phone and user behaviors, and the propagation parameters $n$ and $r_b$ are both unknown and potentially vary with azimuth $\phi$.

However, we do not seek a specific value of $r_{max}$, but rather the minimum distance that is equal to or greater than all plausible values of $r_{max}$, and is therefore also equal to or greater than the maximum plausible distance at which a hand off would occur.
With this in mind, consider the quantity:
\begin{equation}
\frac{r_{max}}{G_T^{1/n}} = 
\left(\frac{\lambda}{4\pi}\right)^{2/n} 
                \left(\frac{P_T G_R}{P_{R,min}} \right)^{1/n} 
                r_b^{1-2/n} 
\label{eRDCL2}
\end{equation}
This quantity can be interpreted as a random variable.
Furthermore, this quantity depends on $\phi$ only to the extent that $n$ and $r_b$ vary with $\phi$.
Since $r_b \propto h_{MS}$, 
and $h_{MS}$ can be treated as a random variable which is independent of $\phi$, so too can $r_b$. 
This leaves $n$, which in principle is both deterministic and variable with $\phi$.   
To make progress, let us treat $n$ as being independent of $\phi$ for now, and then (in later sections) consider the RPL shapes resulting from different fixed values of $n$.  In this way we can assess the impact of varying $n$ with $\phi$ without \emph{a priori} knowledge of how $n$ varies with $\phi$.

These considerations motivate the following choice for the RPL shape: 
\begin{equation}
r_{rpl}(\phi) \propto G_T ^{1/n}(\phi)
\label{eRPLpropto}
\end{equation}
where $r_{rpl}$ is the distance to the RPL boundary.
(It should be noted that $G_T$ also depends on elevation angle; however this angle is approximately equal to the horizon value as long as $r_{max} \gg h_{BS}$.)
This RPL shape has the desired properties that it accounts for the relevant physics (namely, the Friis Equation with a justifiable path loss model), and is independent of hand off considerations.

To be clear, the shape of the tightest possible (i.e., minimum area) RPL depends on the location of nearby BSs; in particular those within the azimuthal span of the sector.  
In practice, the difficulties mentioned previously make it impossible to develop a shape that accounts for the locations of nearby BSs; and in any event (and as explained in Section~\ref{sIntro}) this approach risks conflict with precedent established in U.S. courts.
The shape described by Equation~\ref{eRPLpropto}, scaled by some sufficiently large coefficient, is arguably the best one can do under these constraints. 
This coefficient is smaller than the largest plausible value of the right side of Equation~\ref{eRDCL2}; however this value cannot be determined with suitable accuracy given the available information.
This coefficient must depend on the locations of nearby BSs, but at the same time must not depend on assumptions about hand off behavior.
This is achieved by setting the coefficient based on the spatial density of BSs within the azimuthal span of the sector, as opposed to the distance to particular BSs in particular directions.  
The simplest way to implement this idea is simply to use a weighted mean of the distances to the $M$ nearest BSs lying within the azimuthal half-power beamwidth (HPBW) of the sector antenna. 
Thus, we propose the following definition of the RPL:
\begin{equation}
r_{rpl}(\phi) = c\left[ \frac{1}{M}\sum_{m=1}^M{d_m} \right] \left[ \frac{G_T(\phi)}{G_{T,max}} \right]^{1/n}
\label{eRPL}
\end{equation}
where the second factor is the mean of the distances ($d_m$) from the BS of interest to the $M$ closest BSs within the HPBW of $G_T(\phi)$, 
$G_{T,max}$ is the maximum value of $G_T(\phi)$ (serving to normalize the antenna pattern to a maximum value of 1), 
and $c$ is a constant which for the moment will be set to 1, but which can be adjusted for reasons to be addressed later in this paper.

There are two additional considerations.
First, $r_{rpl}$ must be less than the radio horizon distance $r_h$; i.e., the maximum distance for a terrestrial radio link as determined by the curvature of the Earth. This distance is approximately (see e.g. \cite{RSE-R1E})
\begin{equation}
r_h	\approx \left(4.12~\mbox{km}\right) \sqrt{ \frac{h_{BS}}{\mbox{1 m}} }
\label{eRh}
\end{equation}
Thus, $r_{rpl}(\phi)$ is the lesser of $r_h$ and the value computed from Equation~\ref{eRPL}.
For the experiment reported in Section~\ref{sFE}, BS antenna heights are on the order of 10s of meters, so the corresponding radio horizon is at least 13~km distant.  This is greater than the maximum value of $r_{rpl}$ for all sectors considered.

Second, the path loss model used here does not account for atmospheric absorption.  This is typically not an issue for frequencies below about 10~GHz.
However, to the extent that absorption is significant -- and especially in the millimeter-wave bands -- the path loss model will be invalid and the size of the RPL will be overestimated (i.e., more conservative).  Modifying the path loss model to account for this condition, and then evaluating the resulting performance in field conditions, is left for future work.  

\section{RPL Shape Model}
\label{sRSM}

\begin{figure}
\begin{center}
\includegraphics[width=0.9\columnwidth]{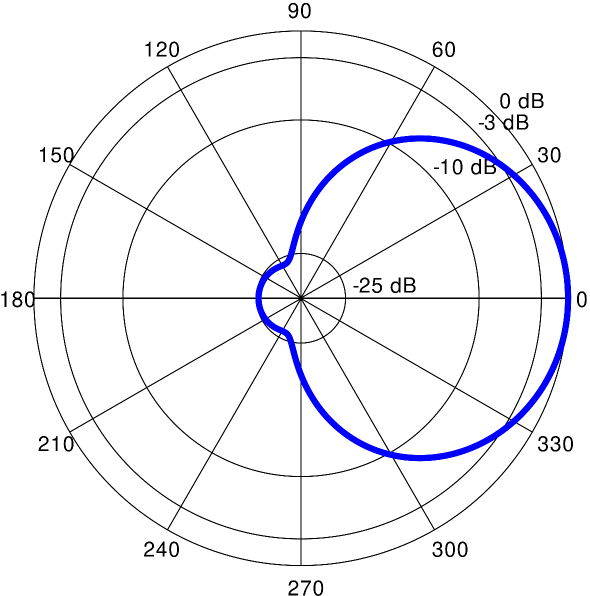}
\end{center}
\caption{\label{fGT}Azimuthal pattern $G_T(\phi)$ according to Equation~\ref{eRPLshape} with $q=6.5$, $p=0.003$, and $\phi_0=0$.}
\end{figure}

As indicated in Equation~\ref{eRPL}, $r_{rpl}(\phi)$ depends on $G_T(\phi)$.
However, in a typical CDR-based localization problem, knowledge of $G_T(\phi)$ is limited.  
It is typical that neither the function $G_T(\phi)$, pattern measurement data, nor the manufacturer and model of the antenna are available.  
Information about $G_T(\phi)$ may be limited to the azimuthal orientation $\phi_0$, HPBW, and perhaps the front-to-back ratio (F/B).
To overcome this problem, we adopt the well-known ``$\cos^q$'' antenna pattern model (see e.g. \cite{ST13}) with a modification to introduce a backlobe:
\begin{equation}
G_T(\phi) = \cos^q\left(\frac{\pi}{2}\sin\frac{\phi-\phi_0}{2}\right) + p \left|\sin\frac{\phi-\phi_0}{2}\right|
\label{eRPLshape}
\end{equation}
where
$q$ sets HPBW, and
$p$ sets F/B.
For example, $q=6.5$ results in HPBW of $66^{\circ}$ (typical for a traditional macrocell), and $p=0.003$ results in a typical F/B of $25$~dB.
Figure~\ref{fGT} shows $G_T(\phi)$ for these values. 

\begin{figure}
\begin{center}
\includegraphics[width=0.9\columnwidth]{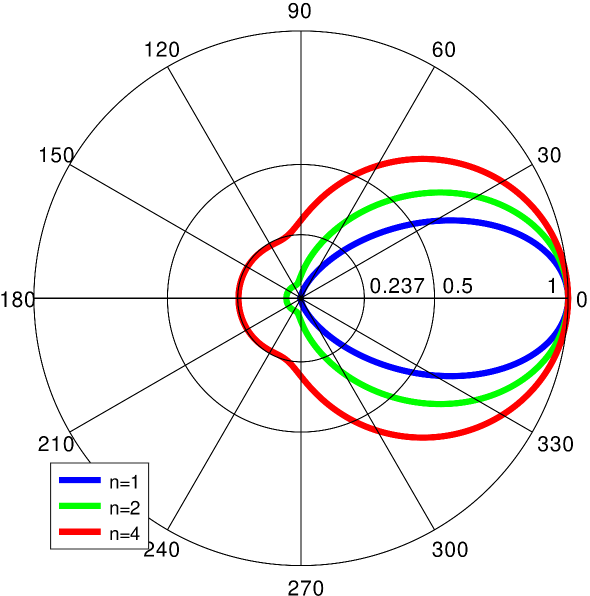}
\end{center}
\caption{\label{fRPLshape}RPL shapes associated with the pattern in Figure~\ref{fGT} with $n=1$ (not physical but included for comparison), $n=2$ (free space), and $n=4$ (typical condition beyond the breakpoint distance).  The radial coordinate is in normalized distance units.}
\end{figure}

The associated RPL shape $G_T^{1/n}(\phi)$ is shown in Figure~\ref{fRPLshape}, for various values of $n$.
Note that the backlobe is relatively unimportant for $n=2$, but contributes significant area to the RPL for $n=4$.  
This is useful information since MSs located in the angular region associated with the backlobe are likely to be closer than the breakpoint distance $r_b$, so it is likely that $n$ is closer to 2 than 4 in these directions. 
Indeed, this is consistent with the results of the experiment documented in Section~\ref{sFE}, where it is found that $n=2$ yields a tighter RPL for MSs in the back half-space $\left|\phi-\phi_0\right| \ge \pi/2$.   

\section{Field Experiment}
\label{sFE}

\begin{figure}
\begin{center}
\includegraphics[width=0.95\columnwidth]{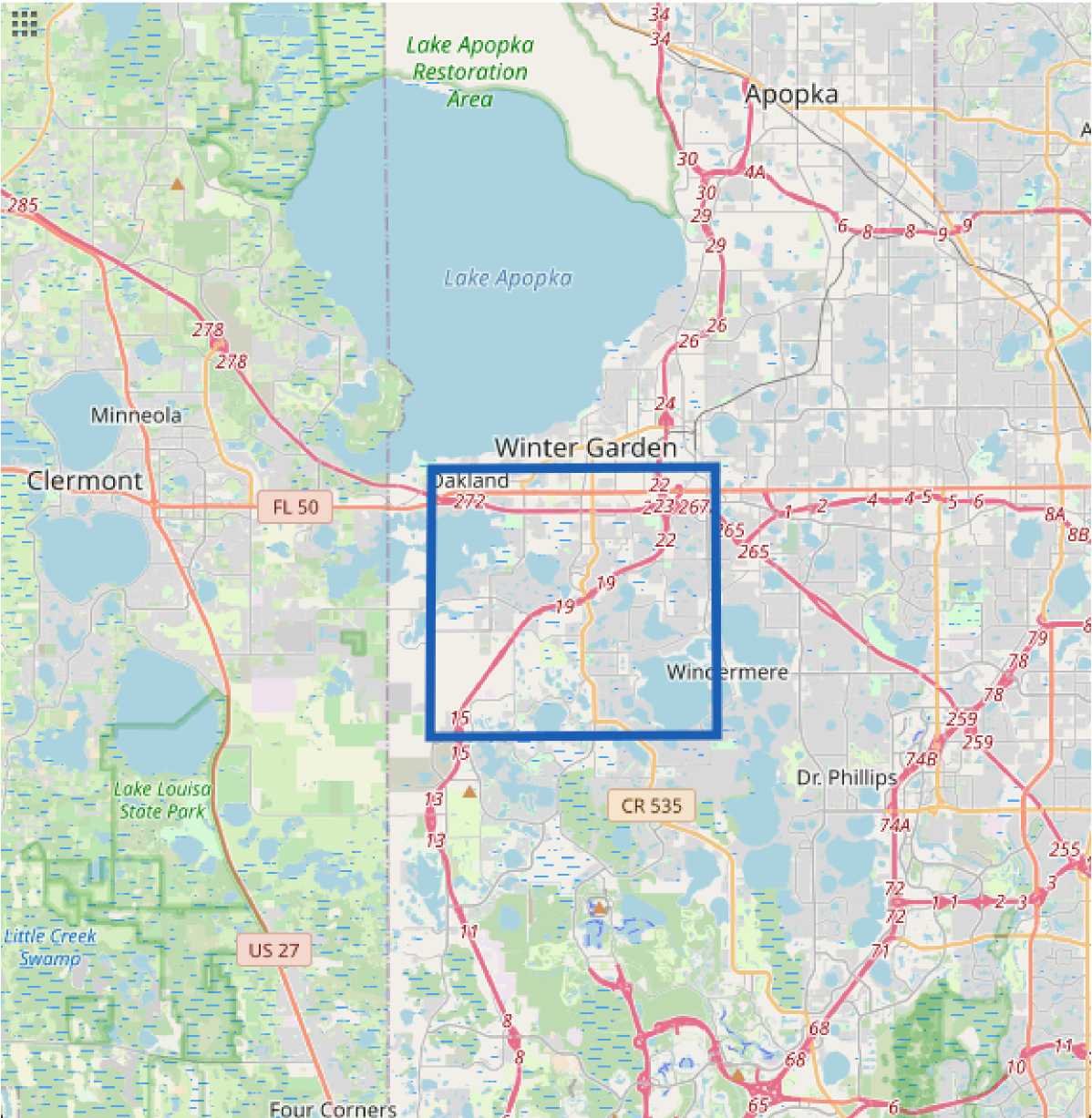}
\end{center}
\caption{\label{fOSM}Region of field testing.  North is up, East is right.  Area shown is about 40~km $\times$ 40~km.  The blue 10~km $\times$ 10~km square bounds the drive test route. Image credit: OpenStreetMap (openstreetmap.org).}
\end{figure}

An experiment was conducted using measurements from a drive test conducted in the vicinity of Winter Garden, Florida, shown in Figure~\ref{fOSM}.  
The region consists of suburbs of Orlando (just beyond the east edge of Figure~\ref{fOSM}) and is essentially flat with negligible variation in elevation.
Data analysis was limited to 700~MHz cellular networks operated by the three primary commercial providers in this area, identified henceforth as Networks~A, B, and C.\footnote{Data presented in this paper is solely for the purpose of evaluating the localization method, and must not be construed as an assessment of network performance. To prevent any misunderstanding in this regard, the identities of the providers are not disclosed.} 

\subsection{Methodology}
\label{ssMethodology}

Commercially-available test equipment
(Rohde \& Schwarz R\&S\textsuperscript{\textregistered}TSMA6B and R\&S\textsuperscript{\textregistered}TSME6, using in-vehicle antennas)
was used to continuously scan for active sectors 
(in LTE terminology, unique combinations of ``eNodeB ID'' and ``physical cell ID'') 
and record ``reference signal received power'' (RSRP) for each.
RSRP is average power received in all LTE resource elements associated with a physical cell ID, and so requires partial decoding of the LTE downlink. 
RSRP is the relevant quantity for the present study since (1) it is not significantly affected by cochannel interference and noise, and (2) LTE MSs uplink this value to the network, where it is subsequently used as the primary metric for determining assignment of the MS to a sector.
Thus, the physical cell ID yielding the largest RSRP value, as measured by the scanner at a particular location, can reasonably be assumed to be the best server for that network at that particular location. 

The association of a measurement location to a sector is made on the basis of best server as described in the previous paragraph.
Note that no connections are made.
This is a limitation since (as noted in Section~\ref{sIntro}) networks may establish and hand off connections based on considerations other than received power.
Unfortunately, it is not practical to actually make connections and track the subsequent hand off behavior.
This is primarily because doing so would require access to information that network operators are not equipped to provide, or are not equipped to provide in the quantities required for this study, or are unwilling to provide due to potential liability, potential for disclosure of technical information to competitors, or legal concerns. 
As explained in \cite{JC22}, best server analysis is regarded by both practitioners and U.S. courts to be a reasonable alternative, with the additional benefit that all networks covering the drive test area may be evaluated simultaneously and in a reproducible manner. 
Thus, the method used in this experiment is appropriate for evaluating the analysis of Section~\ref{sREA}; i.e., Are observed power measurements consistent with the RPLs obtained using the method proposed in Sections~\ref{sREA} and \ref{sRSM}?

\subsection{Data Collection}

\begin{figure}
\begin{center}
\includegraphics[width=0.8\columnwidth]{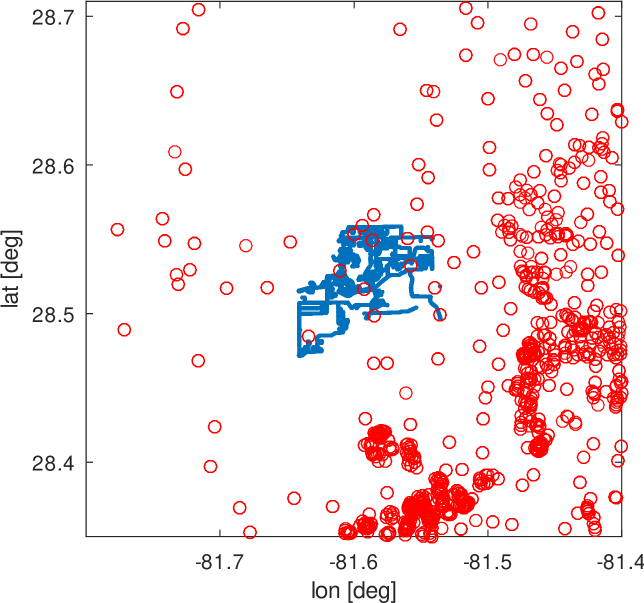}\\ \vspace{0.2in} 
\includegraphics[width=0.8\columnwidth]{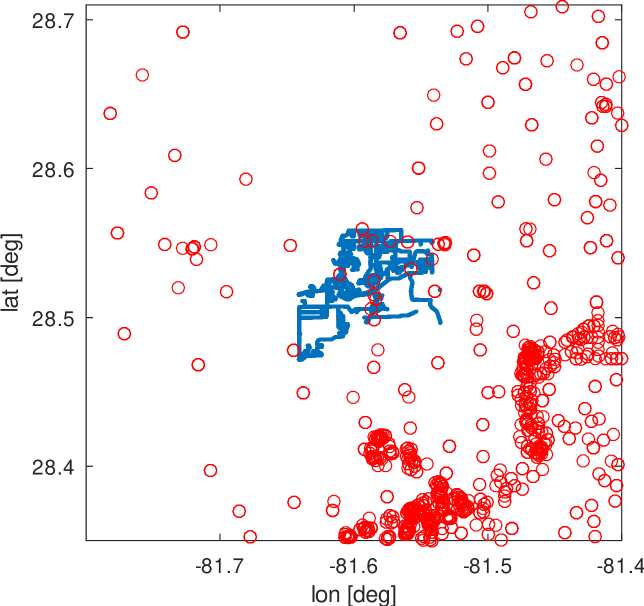}\\ \vspace{0.2in} 
\includegraphics[width=0.8\columnwidth]{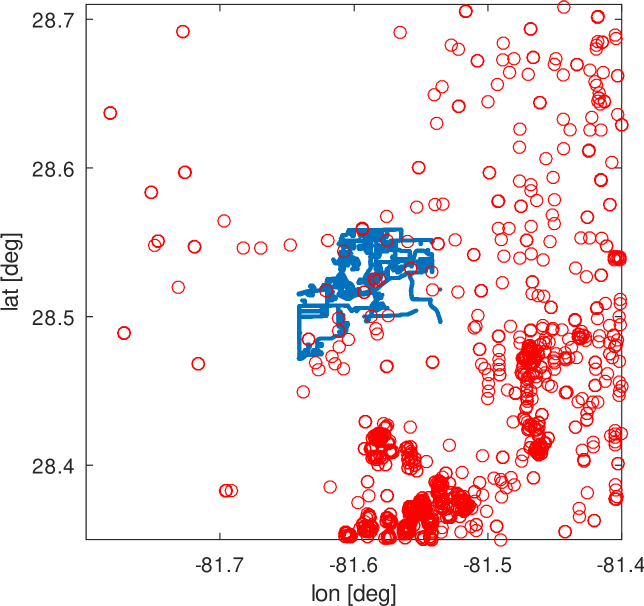}
\end{center}
\caption{\label{fArea}
Region of field testing, scaled to same area as Figure~\ref{fOSM}. 
\emph{Blue:} Drive test route.
\emph{Red:} BSs of Networks A, B, and C (top to bottom, respectively).
}
\end{figure}

\begin{figure}
\begin{center}
\includegraphics[width=0.8\columnwidth]{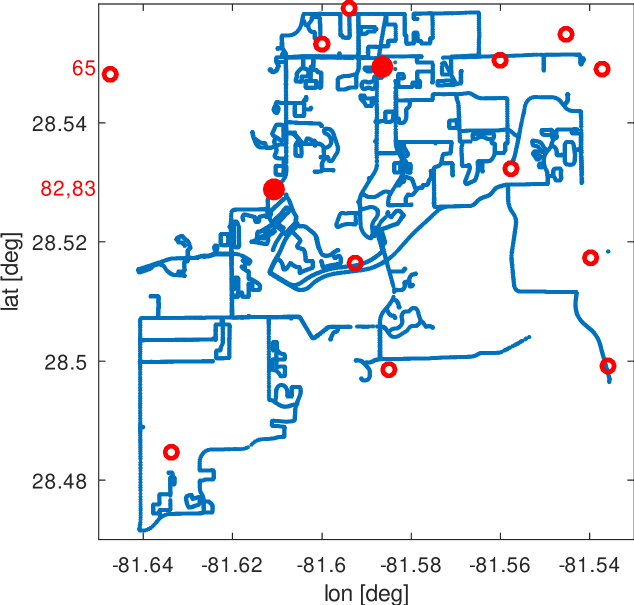}\\ \vspace{0.2in}
\includegraphics[width=0.8\columnwidth]{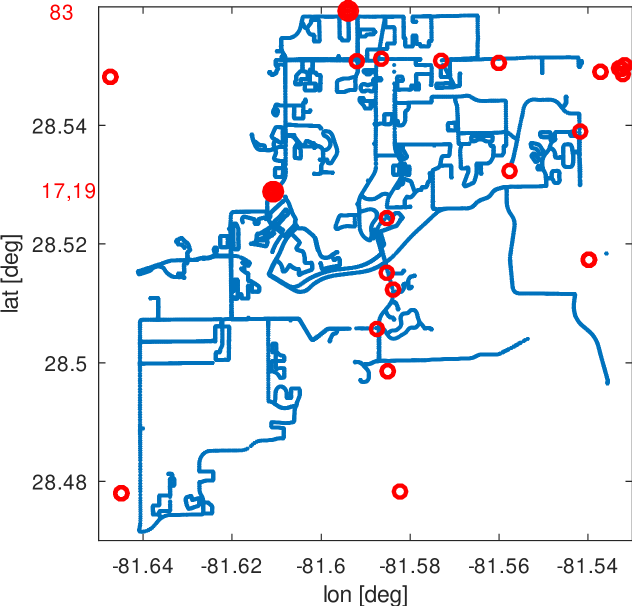}\\ \vspace{0.2in}
\includegraphics[width=0.8\columnwidth]{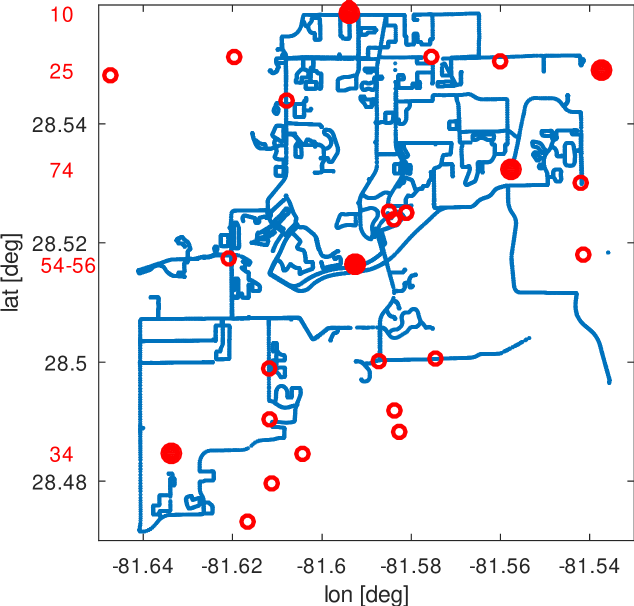}
\end{center}
\caption{\label{fDT}
Same as Figure~\ref{fArea}, zoomed in to the drive test area. Red numbers along the left identify sectors of the BSs represented by filled red circles directly to the right.
}
\end{figure}

\begin{figure}
\begin{center} \includegraphics[width=0.9\columnwidth]{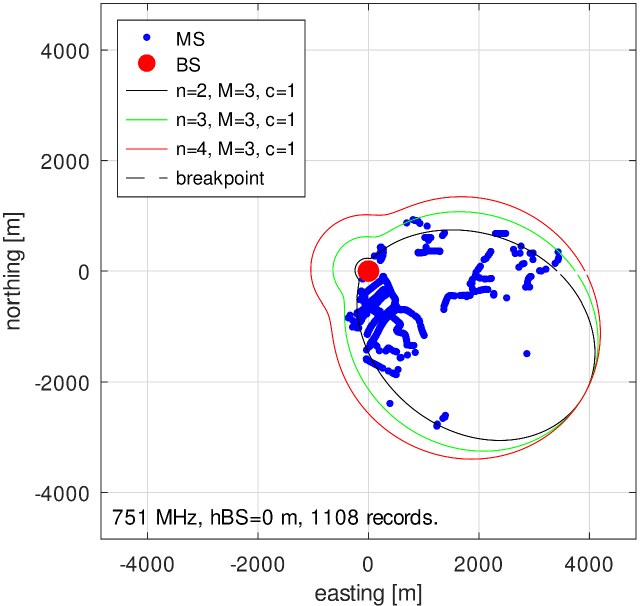} \end{center}
\caption{\label{fB017} Network~B Sector 17.  $n=4$ error rate: 0.0\%. Antenna height is not provided by this operator, so breakpoint distance cannot be calculated.}
\end{figure}

\begin{figure}
\begin{center} \includegraphics[width=0.9\columnwidth]{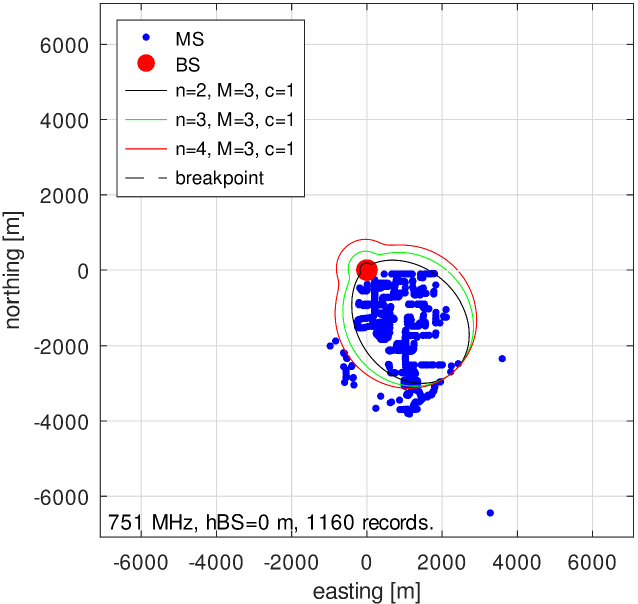} \end{center}
\caption{\label{fB083} Network~B Sector 83.  $n=4$ error rate: 7.67\%. Antenna height is not provided by this operator, so breakpoint distance cannot be calculated.}
\end{figure}

Figure~\ref{fArea} shows the drive test route superimposed on a map of BS locations for each of the three major commercial cellular networks in the area. 
Figure~\ref{fDT} shows the same information, but zoomed in to more clearly show the locations of BSs with respect to the drive test route.
Each blue dot in Figures~\ref{fArea}--\ref{fC010} represent a single record from this dataset.
A total of 49,139 measurement records were acquired from signals received from a total 386 sectors.  

Separately, we obtained infrastructure information for each of the networks.
For each sector, this included
BS location, $\phi_0$, and azimuthal HPBW.
We also obtained antenna heights for two of the three networks.

\subsection{Analysis Procedure}
\label{ssAP}

First, drive test data were merged with the infrastructure data in order to obtain a lists of measurement records for each sector.
For Networks~A and B, we analyzed the results for the three sectors having the largest number of associated measurement records (Sections~\ref{ssRA} and \ref{ssRB}, respectively).
For Network~C
(Section~\ref{ssRC}), we analyzed the results for 10 sectors having the largest number of associated measurement records; this was to include two sectors for which the performance of the RPL estimation method was found to be relatively poor. 

For each sector analyzed, the result is a plot of the locations of measurement records associated with the sector, along with RPLs computed using Equation~\ref{eRPL} with the antenna model of Equation~\ref{eRPLshape} ($q$ set to match HPBW, $p=0.003$) with $M=3$ and $c=1$, while varying $n$ from 2 to 4.
Sector HPBWs were $66^{\circ}\pm 1^{\circ}$ for 12 sectors, and $66^{\circ}\pm 3^{\circ}$ for the remaining 4 sectors; thus,  Figures~\ref{fGT} and \ref{fRPLshape}, as shown, are very close to the actual shapes used. 

For those networks for which we had antenna heights  
(Networks A and C),
we also calculated the breakpoint distance $r_b$ for $h_{MS}=1$~m and included this in the associated figure.
This circle provides some idea of where the path loss exponent is expected to transition from 2 to a higher value.

\subsection{Results: Network~A}
\label{ssRA}

\begin{figure}
\begin{center} \includegraphics[width=0.9\columnwidth]{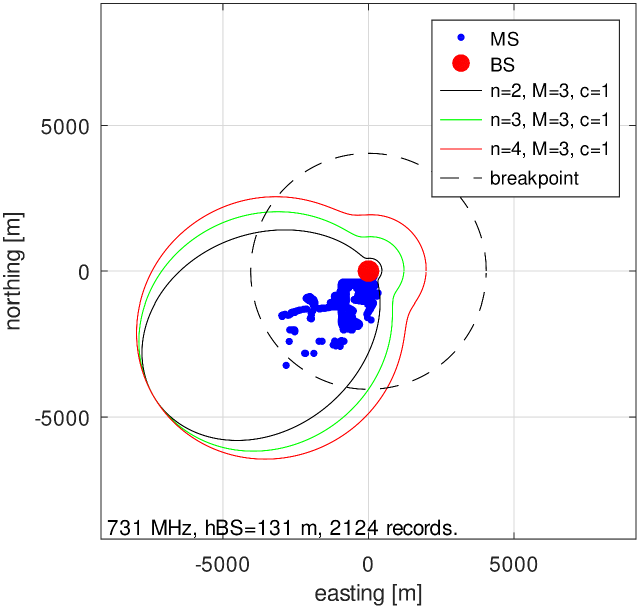} \end{center}
\caption{\label{fA082} Network~A Sector 82.  $n=4$ error rate: 0\%. }
\end{figure}

\begin{figure}
\begin{center} \includegraphics[width=0.9\columnwidth]{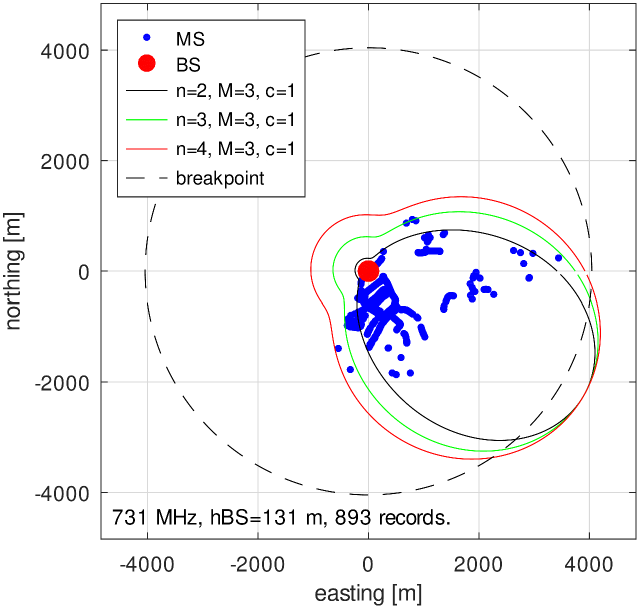} \end{center}
\caption{\label{fA083} Network~A Sector 83.  $n=4$ error rate: 0\%. }
\end{figure}

\begin{figure}
\begin{center} \includegraphics[width=0.9\columnwidth]{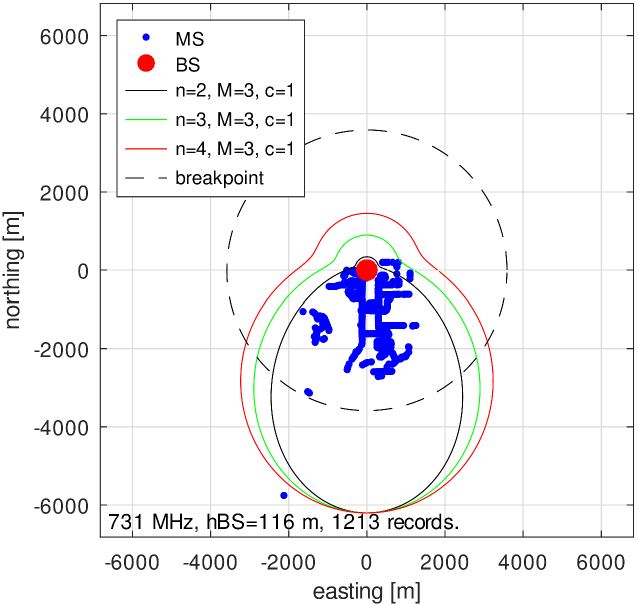} \end{center}
\caption{\label{fA065} Network~A Sector 65.  $n=4$ error rate: 0.08\%. }
\end{figure}

Results for Network~A 
Sectors~82, 83, and 65 are shown in Figures~\ref{fA082}--\ref{fA065}, respectively.
(Note that sector identification numbers are assigned sequentially as part of the study, and do not correspond to ``cell ID'' or other labels used by the network operators.)
Beginning with Sector~82, we see that 100\% of the 2124 associated measurement records fall within the proposed ($n=4$) RPL.
In fact, 100\% of records fall within the $n=2$ RPL: This is not surprising since nearly all measurement records lie within the $h_{MS}=1$ breakpoint distance. 
MS locations for Sectors~83 (adjacent to Sector~82 on the same BS) and 65 also lie within the breakpoint distance, however we see that $n=4$ is required for best performance if the same values of $n$, $M$, and $c$ are to be used across all sectors.

The RPLs obtained for these sectors are ``loose''; that is, one could in principle obtain a smaller RPL (perhaps by using a smaller value of $c$) that nevertheless bounds the MS locations.
However, it is interesting to note that that RPL shape is loose primarily within the antenna HPBW.
Outside this angular span, the RPL is well-matched (``tight'') to the observed distribution of MS locations.
Finally, we note that no MS associations appear in the back lobe of the sector antenna pattern.  This is true also for the remaining sectors in all three networks, suggesting that MS associations through the backlobe of the sector antenna are rare.

\subsection{Results: Network~B}
\label{ssRB}

\begin{figure}
\begin{center} \includegraphics[width=0.9\columnwidth]{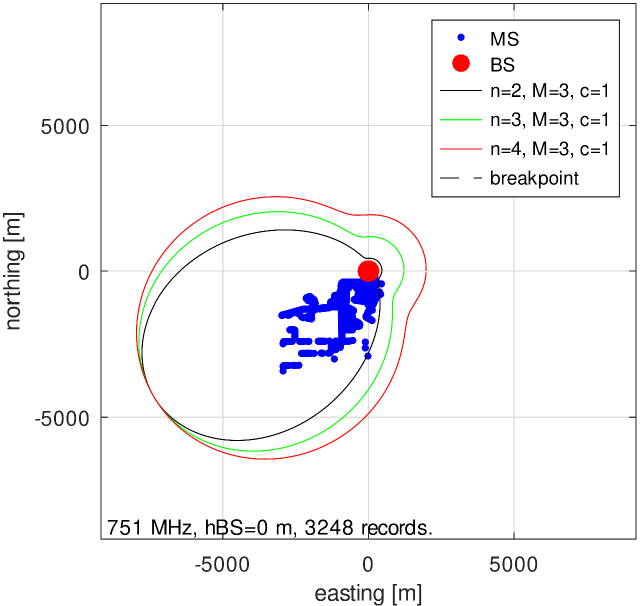} \end{center}
\caption{\label{fB019} Network~B Sector 19.  $n=4$ error rate: 0.0\%. Antenna height is not provided by this operator, so breakpoint distance cannot be calculated.}
\end{figure}

Results for Network~B Sectors~19, 17, and 83 are shown in Figures~\ref{fB019}--\ref{fA083}, respectively.
(It is a coincidence that both Network~B Sector~83 and Network~A Sector~83 were among the top three sectors for these networks.)
Note that Sectors~19 and 17 are adjacent sectors of the same BS.
The results are qualitatively similar to those of Section~\ref{ssRA} with the exception that we see that the RPLs for Sector~83 is ``overtight'', resulting in a relatively large error rate of 7.67\%.
However, it is apparent that this error rate could be reduced to near zero with a small increase in $c$.

\subsection{Results: Network~C}
\label{ssRC}

\begin{figure}
\begin{center} \includegraphics[width=0.9\columnwidth]{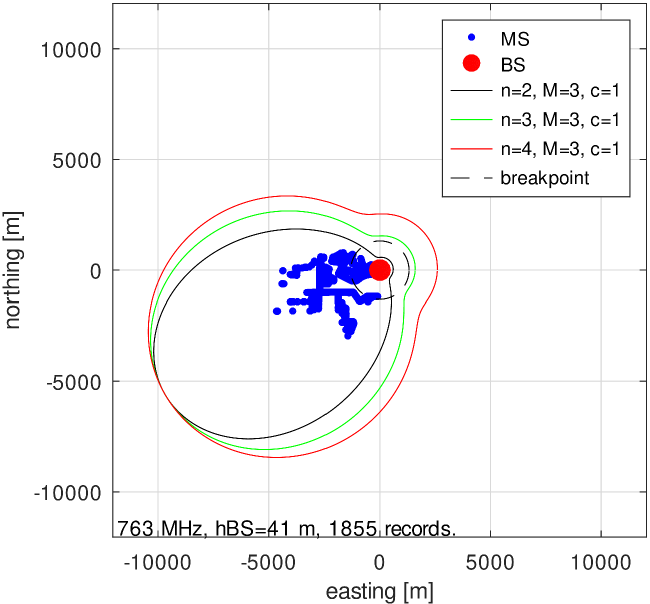} \end{center}
\caption{\label{fC054} Network~C Sector 54.  $n=4$ error rate: 0\%.}
\end{figure}

\begin{figure}
\begin{center} \includegraphics[width=0.9\columnwidth]{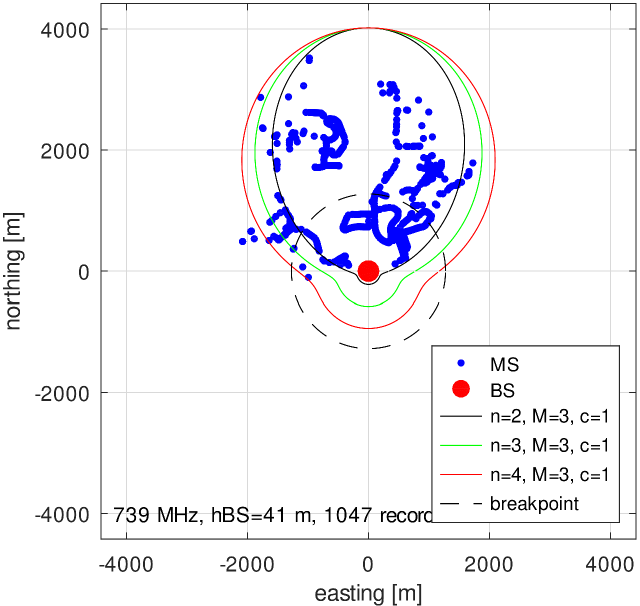} \end{center}
\caption{\label{fC057} Network~C Sector 57.  $n=4$ error rate: 0.38\%.}
\end{figure}

\begin{figure}
\begin{center} \includegraphics[width=0.9\columnwidth]{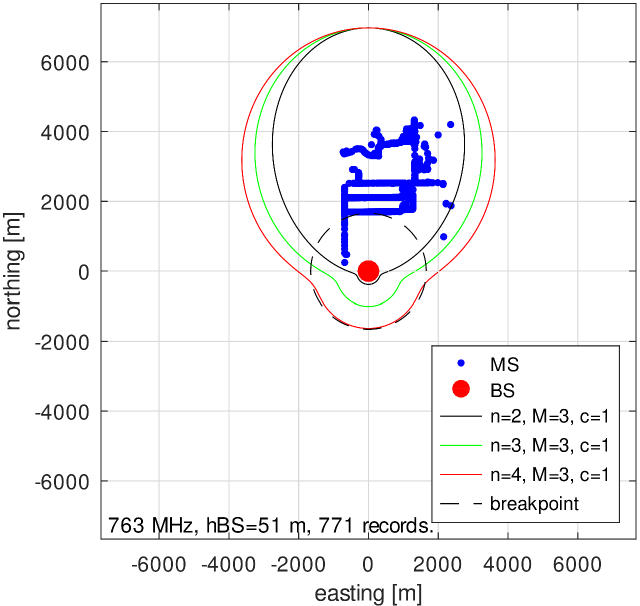} \end{center}
\caption{\label{fC034} Network~C Sector 34.  $n=4$ error rate: 0\%.}
\end{figure}

\begin{figure}
\begin{center} \includegraphics[width=0.9\columnwidth]{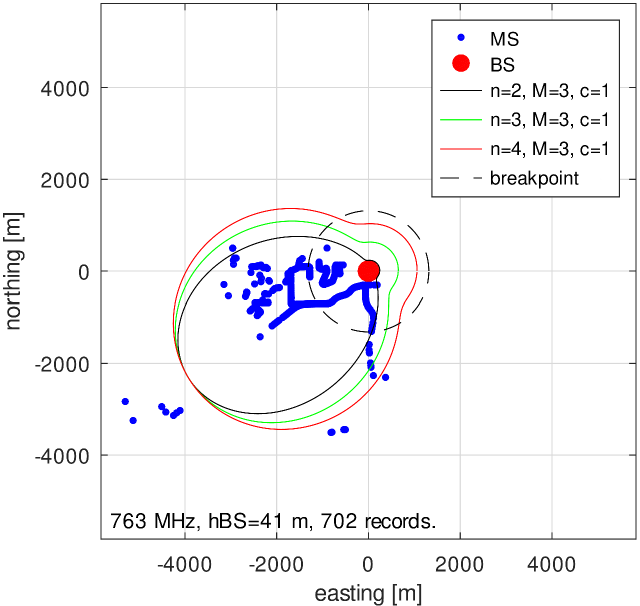} \end{center}
\caption{\label{fC074} Network~C Sector 74.  $n=4$ error rate: 1.71\%.}
\end{figure}

\begin{figure}
\begin{center} \includegraphics[width=0.9\columnwidth]{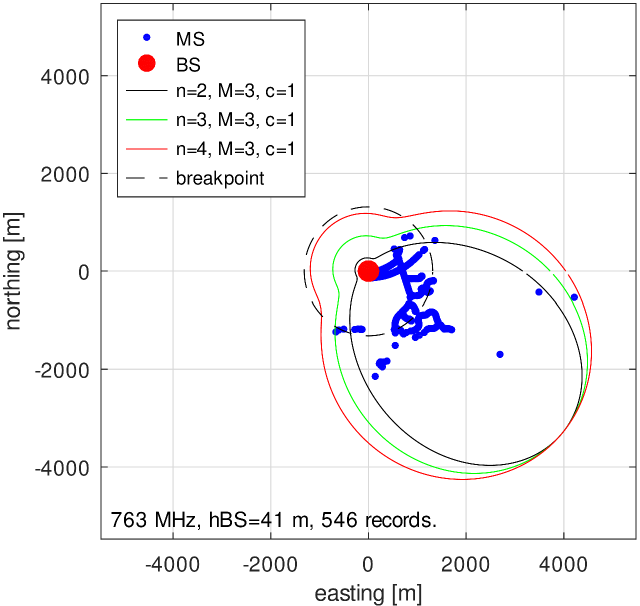} \end{center}
\caption{\label{fC056} Network~C Sector 56.  $n=4$ error rate: 0\%.}
\end{figure}

\begin{figure}
\begin{center} \includegraphics[width=0.9\columnwidth]{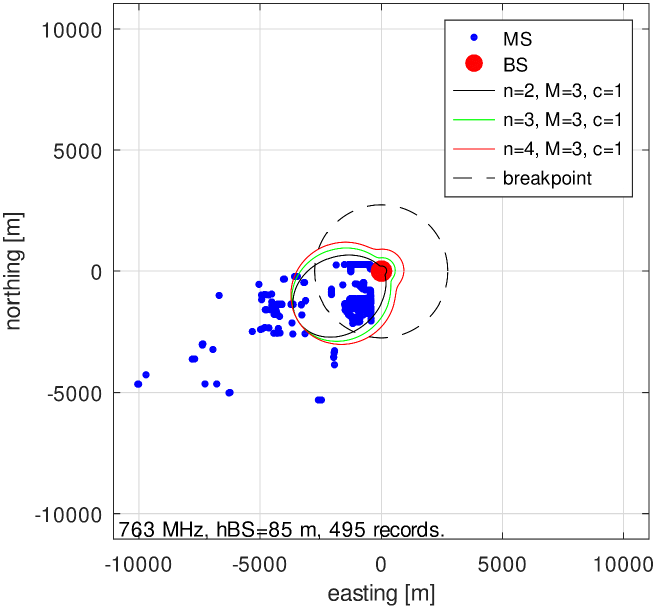} \end{center}
\caption{\label{fC025} Network~C Sector 25.  $n=4$ error rate: 13.1\%.}
\end{figure}

\begin{figure}
\begin{center} \includegraphics[width=0.9\columnwidth]{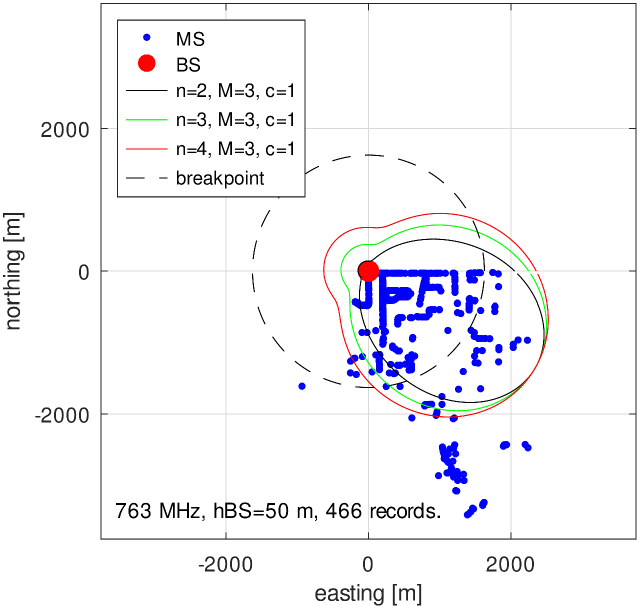} \end{center}
\caption{\label{fC010} Network~C Sector 10.  $n=4$ error rate: 13.3\%.}
\end{figure}

Results for Network~C Sectors~54, 57, 34, 74, 56, 25, and 10 are shown in Figures~\ref{fC054}--\ref{fC010}, respectively.
The results for Sectors~54, 57, 34, 74, and 56 are, again, qualitatively similar to those observed in Sections~\ref{ssRA} and \ref{ssRB}.
The error rate for Sectors~25 and 10, on the other hand, are both relatively high and also would require $c$ to increase to 2.8 in order to reduce the error rate to zero.
The poor performance for these sectors appears to be related to the RPL size calculation, which depends on the distribution of BS locations.
However the problem is not evident from examination of Figure~\ref{fDT}.
We speculate that the problem may be atypical effective isotropic radiated power (EIRP) for the sectors involved in calculations for these two RPLs.

Finally, we note that Network~C Sectors 53, 55, and 39 were also analyzed; however these sectors are identical to Network~C Sectors 54, 57, and 34, respectively, except these sectors used a different frequency in the 700~MHz band. Not surprisingly, the results for the additional sectors are essentially the same and so are not reported here.

\subsection{Summary of Results}
\label{ssSR}

\begin{figure}
\begin{center} \includegraphics[width=0.9\columnwidth]{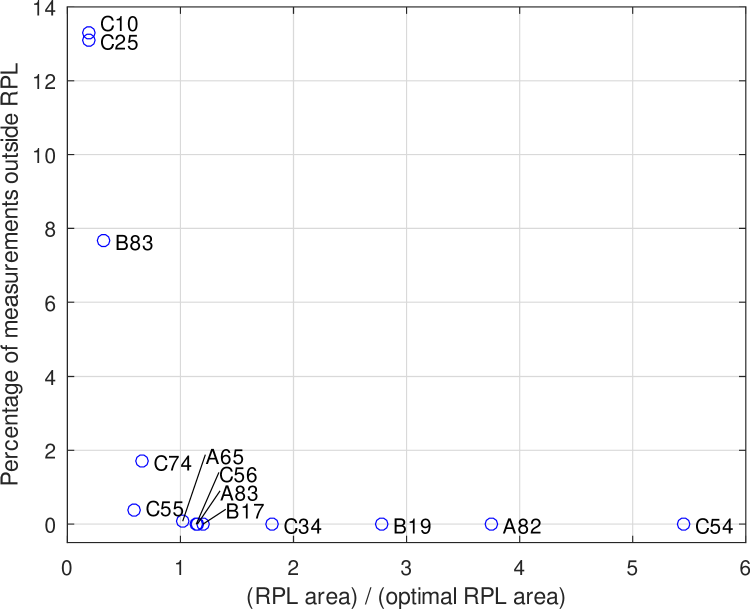} \end{center}
\caption{\label{fSummary} Summary assessment of RPL fit to measurement data.  Markers correspond to the 13 sectors for which results are shown in Sections~\ref{ssRA}, \ref{ssRB}, and \ref{ssRC}.  These are labeled with network name (i.e., A, B, or C) followed by Sector ID.}
\end{figure}

Figure~\ref{fSummary} shows a summary of performance of the RPL estimation method with $n=4$, $M=3$, and $c=1$.
The horizontal axis is the ratio of the area of the RPL to the area of the RPL computed using \emph{optimal} parameters.  
These parameters are determined by repeating the analysis of each sector for $n=2$, $3$, and $4$ as well as $M=1$, $2$, and $3$ (i.e., 9 combinations of $n$ and $M$).  For each combination of $n$ and $M$, $c$ is then determined to be the smallest value that yields an error rate of zero.  The resulting RPL is ``optimal'' in the sense that it yields the smallest RPL area with an error rate of zero.
In Figure~\ref{fSummary}, ideal performance corresponds to an error rate of 0 with an area ratio of 1.
This level of performance is very nearly achieved in Network~A Sectors~65 and 83, Network~B Sector~17, and Network~C Sector~56.  
An error rate of zero is achieved for an additional 4 sectors by being ``loose'' (larger than necessary); these are Network~A Sector~82, Network~B Sector~19, and Network~C Sectors~34 and 54.
Network~B Sector~83 and Network~C Sectors 10, 25, 55, and 74 are ``overtight'', resulting in error rates ranging from 0.08\% to 13.3\%.

It is noted in the previous paragraph that combinations of $n$ and $M$ other then $n=4,M=3$ were examined.
It is found that there is no obviously favored combination of $n$ or $M$ for the optimal RPL.  
It is true that the most popular combination was $n=4,M=3$, which occurred for 3 out of 13 sectors (Network~B Sector~19, Network~C Sector~57 and 10).  The only combination that does not occur is $n=2, M=2$.
In the absence of more or better information, $n=4$, $M=3$ appears to be the best choice. 

A natural question to ask is what is takes to reduce the error rate to zero for all sectors.  Increasing $c$ from 1 to 2.8 (all other parameters fixed) reduces the error rate for all sectors to zero.  Since RPL area increases in proportion to $c^2$, the resulting RPL areas increase by a factor of 7.8, and so most RPLs become very loose.  
This is a fundamental tradeoff in RPL estimation regardless of method. 

Also of interest is the aggregate error rate over all sectors; i.e., the number of measurements whose locations lie outside the estimated RPL, for all sectors combined.
The RPL estimation method with $c=1$ fails for 236 of the 18,471 measurements, yielding an aggregate error rate of less than 1.3\%.
The same calculation done for each network separately yields less than 0.1\% for Network~A (1 out of 4230 measurements) and less than 1.7\% for Network~B (89 out of 5516 measurements) and Network~C (146 out of 8725 measurements).
For Networks~A and B, all of the errors are associated with 1 of 3 sectors.
For Network~C, the error rate is dominated by 2 of 10 sectors. 

\section{Conclusions}
\label{sConc}

This paper presents a method for localization -- specifically, estimation of a ``region of plausible locations'' (RPL) -- that is suitable when the available information is limited to that typically provided in law enforcement and criminal defense applications; namely, the CDR (sector through which the MS is connected), plus static infrastructure information (sector azimuthal orientation, sector azimuthal HPBW, and locations of nearby BSs).  
The RPL is defined by Equation~\ref{eRPL} and (if the sector antenna pattern is not available) Equation~\ref{eRPLshape}.
This method requires values for $n$ (path loss exponent beyond the breakpoint distance), $M$ (number of BSs used to estimate average distance between BSs in the angular span of the sector), and $c$ (RPL size scaling coefficient).
It is shown in the best server field experiment reported in Section~\ref{sFE} (summarized in Figure~\ref{fSummary}) that the method is effective (aggregate error rate 1.3\%) for 700~MHz band cellular networks in flat terrain with parameter values $n=4$, $M=3$, and $c=1$, and that the error rate can be reduced to zero by increasing $c$ to 2.8.
This is an example of the tradeoff between error rate and RPL area that can be expected regardless of the method used to estimate an RPL.

A limitation of the method is that the proposed RPL shape cannot account for persistent deviations from the mean power density due to shadowing and diffraction associated with buildings and terrain. This may explain some fraction of the errors observed in the field experiment reported here, and so this would be an appropriate topic for future work.


\section*{Acknowledgment}

S.W.\ Ellingson was supported in part by a consulting agreement with LexisNexis Risk Solutions. 
The author thanks Trevor Buchanan, Nicholas Butler, Tim DeVries, John Parker, and Kyle Tramonte of the LexisNexis Risk Solutions Special Investigations Unit Geolocation Investigative Team for their helpful review and comments on this work.




\end{document}